\documentclass[journal]{IEEEtran}

\ifCLASSINFOpdf
   \usepackage[pdftex]{graphicx}
   \DeclareGraphicsExtensions{.pdf,.jpeg,.png}
\else
   \usepackage[dvips]{graphicx}
   \DeclareGraphicsExtensions{.eps}
\fi

\usepackage{color}
\usepackage{cite}
\usepackage{amsmath}
\usepackage{bm}
\usepackage{amssymb}
\usepackage{array}
\usepackage{multirow}
\usepackage{makecell}
\usepackage{stfloats}
\usepackage{url}
\usepackage{booktabs}
\usepackage{subfigure}
\usepackage{color}
\usepackage[colorlinks,
            urlcolor=black,
            hyperfootnotes = True,
            citecolor=green]{hyperref}
\usepackage{breakurl}
\newcommand{\PreserveBackslash}[1]{\let\temp=\\#1\let\\=\temp}
\newcolumntype{C}[1]{>{\PreserveBackslash\centering}p{#1}}
\newcolumntype{R}[1]{>{\PreserveBackslash\raggedleft}p{#1}}
\newcolumntype{L}[1]{>{\PreserveBackslash\raggedright}p{#1}}

\begin{document}

\title{A Survey on Rain Removal from \\ Video and Single Image}

\author{Hong Wang, Yichen~Wu, Minghan~Li, Qian~Zhao,
        and~Deyu~Meng,~\IEEEmembership{Member,~IEEE}
\thanks{H. Wang, Y. Wu, M. Li, Q. Zhao, and D. Meng (corresponding author) are with Institute for Information and System Sciences and Ministry of Education Key Lab of Intelligent Networks and Network Security, Xi'an Jiaotong University,
Shaan'xi, 710049 P.R. China}
}



\maketitle

\begin{abstract}
   Rain streaks might severely degenerate the performance of video/image processing tasks. The investigation on rain removal from video or a single image has thus been attracting much research attention in computer vision and pattern recognition, and various methods have been proposed against this task recently. However, there is still not a comprehensive survey paper to summarize current rain removal methods and fairly compare their generalization performance, and especially, still not a off-the-shelf toolkit to accumulate recent representative methods for easy performance comparison and capability evaluation. Aiming at this meaningful task, in this study we present a comprehensive review for current rain removal methods for video and a single image. Specifically, these methods are categorized into model-driven and data-driven approaches, and more elaborate branches of each approach are further introduced. Intrinsic capabilities, especially generalization, of representative state-of-the-art methods of each approach have been evaluated and analyzed by experiments implemented on synthetic and real data both visually and quantitatively. Furthermore, we release a comprehensive repository, including direct links to 74 rain removal papers, source codes of 9 methods for video rain removal and 20 ones for single image rain removal, 19 related project pages, 6 synthetic datasets and 4 real ones, and 4 commonly used image quality metrics, to facilitate reproduction and performance comparison of current existing methods for general users. Some limitations and research issues worthy to be further investigated have also been discussed for future research of this direction.
\end{abstract}

\begin{IEEEkeywords}
    Rain removal, maximum a posterior estimation, deep learning, generalization.
\end{IEEEkeywords}

\section{Introduction}
\IEEEPARstart{I}{mages} and videos captured from outdoor vision systems are often affected by rain. Specifically, as a complicated atmospheric process, rain can cause different types of visibility degradations. Typically, nearby rain drops/streaks incline to obstruct or distort background scene contents and distant rain streaks tend to generate atmospheric veiling effects like mist or fog and blur the image contents~\cite{jorder, tpami,benchmark}. Rain removal has thus become a necessary preprocessing step for subsequent tasks, like object tracking~\cite{5}, scene analysis~\cite{7}, person reidentification~\cite{pid}, and event detection~\cite{2}, to further enhance their performance. Therefore, as an important research topic, removing rain streaks from videos and images has been attracting much attention recently in the filed of computer vision and pattern recognition~\cite{ gk_vision, zhang,dsc,lp,gu}.

In the recent years, various methods have been proposed for this rain removal task for both video and single image~\cite{clear,did,ddn,rescan,wei,csc,liujia,liujia2}. Comparatively, removing rain from an individual image is evidently more challenging than that from a video composing of a sequence of image frames, due to the lack of beneficial temporal information in the former case~\cite{pre,spa,pan}. The methodologies designed for two cases are thus with significant distinction. Yet similarly for both issues, conventional methods mainly adopt the model-driven methodology and especially focus on sufficiently utilizing and encoding physical properties of rain and prior knowledge of background scenes into an optimization problem and designing rational algorithms to solve it, while more recently raised methods often employ the data-driven manner by designing specific network architectures and pre-collecting rainy-clean image pairs to learn network parameters to attain complex rain removal functions \cite{semi,liu}. Most of these methods have targeted certain insightful aspects of the rain removal issue and have their suitability and superiority on specific occasions.

Albeit raising so many methods for rain removal from both video and single image, to the best of our knowledge, there is still not a comprehensive survey paper to specifically summarize and categorize current developments along this research line. Especially, there does still not exist a easily-usable source which could provide an off-the-shelf platform for general users to attain the source codes of current methods presented along this research line for easy performance comparison and capability evaluation for these methods. This, however, should be very meaningful for further prompting the frontier of this research issue and for facilitating an easy performance reproduction of previous algorithms and discovery of more intrinsic problems existing in current methods.

\begin{figure}[htb]
    \vspace{-8mm}
    \hspace{-6mm}
      \includegraphics[scale=0.365]{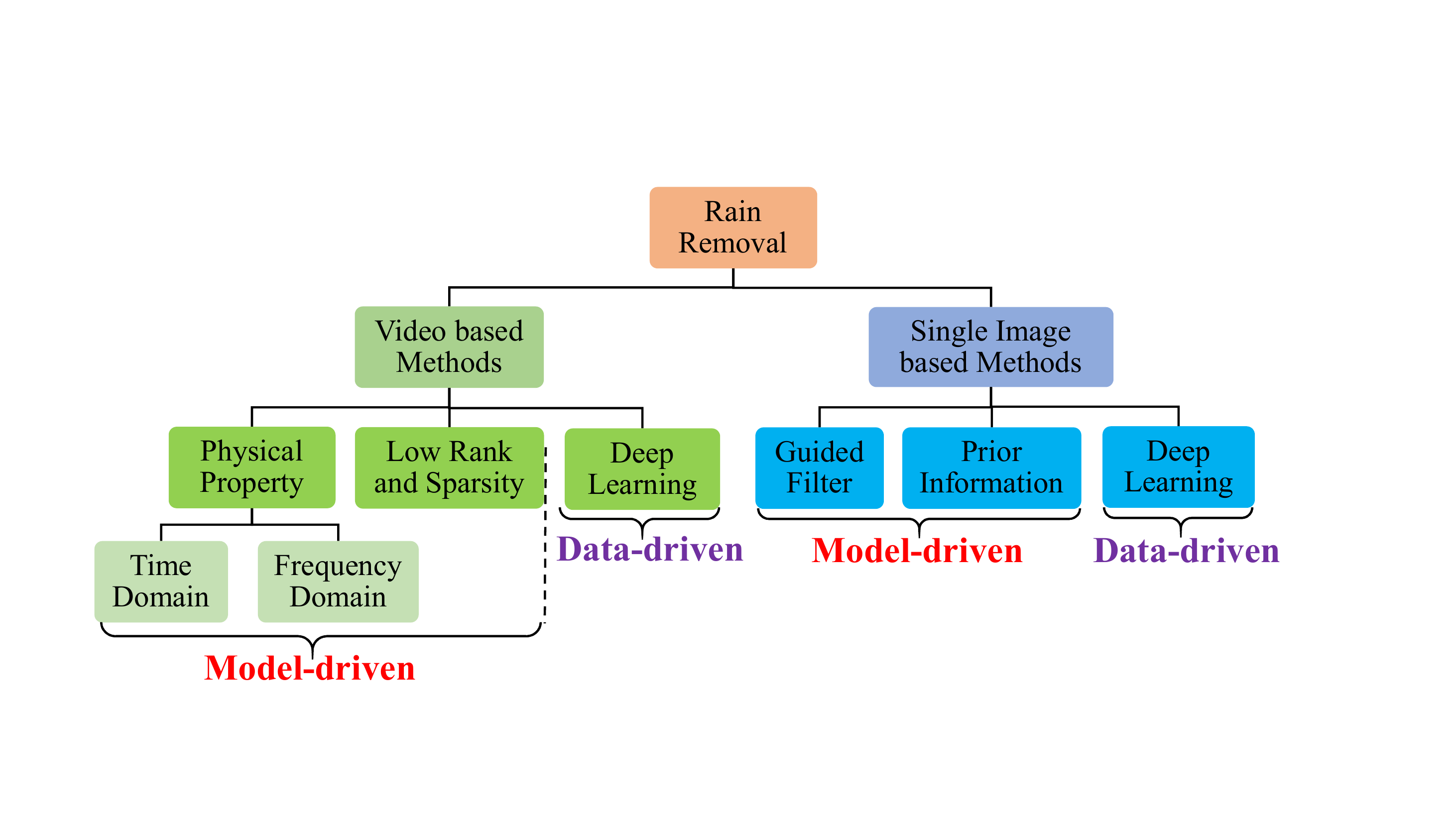}
        \vspace*{-17mm}
    \caption{A hierarchical categorization of current rain removal methods for video and single image.}\vspace{-5mm}
    \label{frame}
\end{figure}

Against this meaningful task, in this study we aim at presenting a possibly comprehensive review for current rain removal methods for video and single image, as well as evaluating and analyzing the intrinsic capabilities, especially generalization, of representative state-of-the-art methods. In all, our contributions can be mainly summarized as follows:

Firstly, we comprehensively introduce the main ideas of the current rain removal methods for both video and single image. In specific, we summarize the physical properties of rain commonly used for rain modeling by previous research. For video and single image rain removal methods raised in both conventional model-driven and latest data-driven manners, we elaborately categorize them into several hierarchal branches, as shown in Fig. \ref{frame}, and introduce the main methodology and representative methods of each branch.

Secondly, we provide a comprehensive performance comparison on representative rain removal methods and evaluate their respective capacity, especially generalization capability, both visually and quantitatively, based on  typical synthetic and real datasets containing diverse rain configurations. The implemented deraining methods, including 7 ones for video and 10 ones for single image, cover recent state-of-the-art model-driven and data-driven rain removal algorithms.

Most importantly, in this study we release a comprehensive repository to facilitate an easy use and performance reproduction/comparison of current rain removal methods for general users. Particularly, this repository includes direct links to 74 rain removal papers, source codes of 9 methods for video deraining and 20 ones for single image deraining, 19 related project pages, 6 synthetic training datasets and 4 real ones, and 4 commonly used image quality metrics.

The rest of the paper is organized as follows. Section \uppercase\expandafter{\romannumeral2} surveys the main contents of recent literatures raised on rain removal issue. Experiments are then presented in Sections \uppercase\expandafter{\romannumeral3} for performance evaluation. Section IV concludes the whole paper, and lists some limitations and research issues worthy to be further investigated for future research of this direction.

\section{Review of Current Rain Removal Methods}
In this section, we first introduce some physical properties of rain, which constitute the modeling foundation of most rain removal methods, and then review the deraining methods for video and single image, respectively, according to the categorization as displayed in Fig. \ref{frame}.

\subsection{Physical Properties of Rain}
A falling raindrop undergoes rapid shape distortions caused by many key factors, such as surface tension, hydrostatic pressure, ambient illumination, and aerodynamic pressure~\cite{bear,tri_review}. These distortions will appear in forms of rain streaks with different brightness/directions~\cite{tri} and distort background objects/scenes of videos/images~\cite{gk_when}. In the following, we introduce some intrinsic properties of rain demonstrated in a video or a single image, which represent the typical clues for optimization or network modeling for constructing a rain removal method.

\subsubsection{\textbf{Geometric Property}}
Beard and Chuang~\cite{bear} described the shape of small raindrops as a sphere, expressed as:
\begin{equation}\label{1}
r\left ( \theta  \right ) =a\left ( 1+\sum_{n=1}^{10} c_{n}\text{cos}\left ( n\theta \right )\right ),
\end{equation}
where $a$ is the radius of the undistorted sphere, $c_{n}$ is the shape coefficient that depends on the radius of raindrop, and $\theta$ is the polar angel of elevation. $\theta=0 $ represents the direction of the rainfall and $r\left ( \theta  \right )$ is the polar radius in the direction of $\theta$.

As a raindrop falls, it attains a constant velocity, called terminal velocity~\cite{foot}. By fitting a large amount of experimental data with the least squares, Foote and Toit \cite{foot} obtained the relationship between the terminal velocity $v$ (m/s) of a raindrop and its diameter $d$ (mm) as:
\begin{equation}\label{2}
\begin{aligned}
&v_{0} = -0.2+5d-0.9d^{2}+0.1d^{3},\\
&v = v_{0}\left ( \rho _{0}/\rho \right )^{0.4},
\end{aligned}
\end{equation}
where $\rho$ is the air density at the location of the raindrop. $\rho _{0}$ and $v_{0}$ are obtained under the 1013 mb atmospheric conditions. Although a strong wind tends to change the rain orientation, the direction of rain streaks captured in the limited range of a video frame or an image is almost consistent~\cite{bossu}.

\subsubsection{\textbf{Brightness Property}}
Garg and Nayar~\cite{gk_vision} pointed out that raindrops can be viewed as optical lens that refract and reflect lights, and when a raindrop is passing through a pixel, the intensity of its image $I_{r}$ is brighter than the background~\cite{gk_photo}. The imaging process was illustrated as:
\begin{equation}\label{3}
I_{r}\left ( x,y \right )=\int_{0}^{\tau }E_{r}\left ( x,y \right )dt+\int_{\tau}^{T }E_{b}\left ( x,y \right )dt,
\end{equation}
where $\tau$ is the time during which a raindrop projects onto the pixel location $(x,y)$ and $T$ is the exposure time of a camera. $E_{r}$ is the irradiance caused by the raindrop and $E_{b}$ is the average irradiance of the background~\cite{gk_vision,liu_pixel}.

%

\subsubsection{\textbf{Chromatic Property}}
Zhang \emph{et al.}~\cite{zhang} made further investigation about the brightness property of rain and showed that the increase in the intensities of R, G, and B channels is dependent on the background scene. By empirical examples, they found that the field of views (FOVs) of red, green, and blue lights are all around $165^{\text{o}}$. For ease of computation, the authors directly assumed that the means of $\Delta R$,  $\Delta G$, and $\Delta B$ are roughly equivalent for pixels covered by raindrops, where $\Delta R$, $\Delta G$, and $\Delta B$ denote the changes in color components of one pixel in two consecutive frames.


\subsubsection{\textbf{Spatial and Temporal Property}}
As raindrops are randomly distributed in space and move at high velocities, they often cause spatial and temporal intensity fluctuations in a video, and a pixel at particular position is not always covered by these raindrops in every frame~\cite{zhang}. Therefore, in a video with stationary scene captured by a stationary camera, the intensity histogram of a pixel sometimes covered by rain exhibits two peaks, one for the background intensity distribution and the other for the rain intensity distribution. However, the intensity histogram of a pixel never covered by rain throughout the entire video exhibits only one peak~\cite{zhang}.

\subsection{Video Rain Removal Methods}
Garg and Nayar \cite{gk_vision, gk_when} made early attempt for rain removal from videos, and proposed that by directly increasing the exposure time or reducing the depth of field of a camera, the effects of rain can be reduced or even removed without altering the appearance of the scene in a video. However, this method fails to deal with heavy rain and fast-moving objects that are close to the camera, and the camcorder setting cannot be adjusted by this method without substantial performance degradation of videos~\cite{tri_review}.

In the past few years, more intrinsic properties of rain streaks have been explored and formulated in algorithm designing for rain removal from videos in static/dynamic scenes. These algorithms can be mainly divided into four categories: time domain based ones, frequency domain based ones, low rank and sparsity based ones, and deep learning based ones. The first three categories follow the hand-crafting pipelines to model rain context and thus should be seen as model-driven methodologies, whereas the latter one follows data-driven manner where features are automatically learnt from pre-collected training data (rainy/clean frame pairs)~\cite{wei,liu}.

\subsubsection{\textbf{Time domain based methods}}
Garg and Nayar~\cite{gk_det} firstly presented a comprehensive analysis of visual effects of rain on an imaging system and then developed a rain detection and removal algorithm for videos, which utilized a space-time correlation model to capture the dynamics of rain and a physics-based motion blur model to explain the photometry of rain. Here the authors assumed that as raindrops fall with some velocity, they affect only a single frame. Hence, the rain streaks can be removed by exploiting the difference between consecutive frames~\cite{tri_video}.

To further improve the rain detection accuracy, Zhang \emph{et al.}~\cite{zhang} incorporated both temporal and chromatic properties of rain and utilized K-means clustering to identify the background and rain streaks from videos. The idea works well in handling light and heavy rain, as well as rain in/out of focus. However, the method often tends to blur images due to a temporal average of the background. To alleviate this problem, Park \emph{et al.}~\cite{park} further proposed to estimate the intensity of pixels and then remove rain recursively by Kalman filter, which performs well in a video with stationary background.

Later, by introducing both optical and physical properties of rain streaks, Brewer \emph{et al.}~\cite{brew} proposed to first identify rain-affected regions showing a short-duration intensity spike, and then replaced the rain-affected pixel with average value in consecutive frames. Naturally, the method is able to distinguish intensity changes caused by rain from those made by scene motion. Yet, it is not very suitable to detect heavy rain where multiple rain streaks overlap and form undesirable shapes.

Zhao \emph{et al.}~\cite{zhao} used temporal and spatial properties of rain streaks to design a histogram model for rain detection and removal, which embedded a concise K-means clustering algorithm with low complexity~\cite{zhang}. To handle both dynamic background and camera motion, Bossu \emph{et al.}~\cite{bossu} utilized a Gaussian mixture model (GMM) and geometric moments to estimate the histogram of orientation of rain steaks.

Inspired by Bayesian theory, Tripathi \emph{et al.}~\cite{tri} relied on temporal property of rain to build a probabilistic model for rain streaks removal. Since intensity variations of rain-affected and rain-free pixels differ by the symmetry of waveform, the authors used two statistical features (intensity fluctuation range and spread asymmetry) for distinguishing rain from rain-free moving object. As there is no any assumption about the shape and size of raindrops, the method is robust to rain conditions. To further reduce the usage of consecutive frames, the authors turned to employing spatio-temporal process~\cite{tri_video}, which has less detection accuracy but better perceptual quality than\cite{tri}.

\subsubsection{\textbf{Frequency domain based methods}}
Barnum \emph{et al.}~\cite{barnum,barnum2} demonstrated a spatio-temporal frequency based method for globally detecting rain
and snow with a physical and statistical model, where the authors utilized a blurred Gaussian model to approximate the blurring effects produced by the raindrops and a frequency-domain filter to reduce the visibility of raindrops/snow. The idea still works in videos with both scene and camera motions and can efficiently analyze repeated rain patterns. Nevertheless, the blurred Gaussian model cannot always cover rain streaks which are not sharp enough. Besides, the frequency-based detection manner often has errors when the frequency components of rain are not in order \cite{tri}.

\subsubsection{\textbf{Low rank and sparsity based methods}}
In the recent decade, low rank and sparsity properties are extensively studied for rain/snow removal from videos. Chen \emph{et al.}~\cite{chen} first considered the similarity and repeatability of rain streaks and generalized a low-rank model from matrix to tensor structure to capture the spatio-temporally correlated rain streaks.
To deal with highly dynamic scenes~\cite{gk_det, park, zhao, tri}, Chen \emph{et al.} further designed an algorithm based on motion segmentation of dynamic scene~\cite{chen2}, which first utilized photometric and chromatic constraints for rain detection and then applied rain removal filters on pixels such that their dynamic property as well as motion occlusion clue were incorporated. Spatial and temporal information is thus adaptively exploited during rain pixel recovery by the method, which, however, still cannot finely fit camera jitters~\cite{ren}.

Later, Kim \emph{et al.}\cite{kim} proposed to subtract temporally warped frames from the current frame to obtain an initial rain map, and then
decomposed it into two types of basis vectors (rain streaks and outliers) via SVM. Next, by finding the rain map to exclude the outliers and executing low rank matrix completion, rain streaks could be removed. Obviously, the method needs extra supervised samples to train SVM.

Considering heavy rain and dynamic scenes, Ren \emph{et al.}\cite{ren} divided rain streaks into sparse and dense layers, and formulated the detection of moving objects and sparse rain streaks as a multi-label Markov random field (MRF) and dense ones as Gaussian distribution.

Jiang \emph{et al.}\cite{jiang,jiang2} proposed a novel tensor based video rain streaks removal approach by fully analyzing the discriminatively intrinsic characteristics of rain streaks and clean videos. In specific, rain streaks are sparse and smooth along the direction of raindrops, and clean videos possess smoothness along the rain-perpendicular direction and global and local correlation along time direction.

Different from previous rain removal methods formulating rain streaks as deterministic message, Wei \emph{et al.}\cite{wei} first encoded the rain layer as a patch-based mixture of Gaussian (P-MoG). By integrating the spatio-temporal smoothness configuration of moving objects and low rank structure of background scene, the authors proposed a concise P-MoG model for rain streaks removal. Such stochastic manner makes the model capable of adapting a wider range of rain variations.

Motivated by the work~\cite{zhangcsc}, Li \emph{et al.}\cite{csc} considered two intrinsic characteristics of rain streaks in videos, i.e., repetitive local patterns sparsely scattered over different positions of a video and multiscale configurations due to their occurrence on positions with different distances to the cameras. The authors specifically formulated such understanding as a multi-scale convolutional sparse coding model (MS-CSC). Similar to \cite{wei}, the authors separately use $L_{1}$ and total variation (TV) to regularize the sparsity of feature maps and the smoothness of moving object layer. Such an encoding manner makes the model capable of properly extracting natural rain streaks from rainy videos.
\subsubsection{\textbf{Deep learning based methods}}
Very recently, deep learning based methods have also been investigated for the video rain removal task. For example, Chen \emph{et al.}\cite{chen3} proposed a convolutional neural network (CNN) framework for video rain streaks removal, which can handle torrential rain fall with opaque streak occlusions. In the work, superpixel has been utilized as the basic processing unit for content alignment and occlusion
removal in a video with highly complex and dynamic scenes.

By exploring the wealth of temporal redundancy in videos, Liu \emph{et al.}~\cite{liujia} built a hybrid rain model to depict both rain streaks and occlusions as:
\begin{equation}\label{liu}
\mathbf{O}_{t}=\left(1-\alpha_{t}\right)\left(\mathbf{B}_{t}+\mathbf{R}_{t}\right)+\alpha_{t} \mathbf{A}_{t}, t=1,2, \ldots, N,
\end{equation}
where $t$ and $N$ signify the current time-step and total number of the frames in a video. $\mathbf{O}_{t}\in \mathbb{R}^{h \times w}$, $\mathbf{B}_{t}\in \mathbb{R}^{h \times w}$, and $\mathbf{R}_{t}\in \mathbb{R}^{h \times w}$ are the rainy image,  background frame, and rain streak frame, respectively. $\mathbf{A}_{t}$ is the rain reliance map and $\alpha_{t}$ is an alpha matting map.
Based on the model (\ref{liu}), the authors utilized a deep recurrent convolutional network (RNN) to design a joint recurrent rain removal and reconstruction network (J4R-Net) that seamlessly integrated rain degradation classification, spatial texture appearances based rain removal, and temporal coherence based background details reconstruction. To address deraining with dynamically detected video contexts, the authors chose a parallel technical route and further developed a dynamic routing residue recurrent network (D3R-Net) as well as a spatial temporal residue learning module for video rain removal~\cite{liujia2}.

\subsection{Single Image Rain Removal Methods}
In contrast to video based deraining methods with temporal redundancy knowledge, removing rain from individual images is more challenging since less information is available. To handle the problem, the algorithm design for single image rain removal has drawn increasingly more research attention. Generally, the existing single image rain removal methods can be divided into three categories: filter based ones, prior based ones, and deep learning based ones.

\subsubsection{\textbf{Filter based methods}}
Xu \emph{et al.}\cite{xu} proposed a single image rain removal algorithm with guided filter~\cite{he_gui}. In specific, by using chromatic property of rain streaks, the authors first obtained coarse rain-free image (guidance image) and then filtered rainy image to get the rain-removed image. For better visual quality, the authors incorporated brightness property of rain streaks and remended the guidance image~\cite{xu2}.

Zheng \emph{et al.}~\cite{zheng} later presented a multiple guided filtering based single image rain/snow removal method. In the work, the rain-removed image was acquired by taking the minimum value of rainy image and the coarse recovery image obtained by merging low frequency part (LFP)  of rainy image with high frequency part (HFP) of rain-free image. To improve the rain removal performance, Ding \emph{et al.}~\cite{ding} designed a guided $L_0$ smoothing filter to get coarse rain-/snow-free image.

Considering that a typical rain streak has an elongated elliptical shape with a vertical orientation, Kim \emph{et al.}~\cite{kim_filter} proposed to detect rain streak regions by analyzing rotation angle and aspect ratio of the elliptical kernel at each pixel, and then executed nonlocal means filtering on the detected regions by adaptively selecting nonlocal neighbor pixels and the corresponding weights.

\subsubsection{\textbf{Prior based Methods}}
In the recent years, maximum a posterior (MAP) has been attracting considerable attention for rain streak removal from a single image~\cite{liubilevel,map}, which can be mathematically described as:
\begin{equation}
\underset{\mathbf{B}, \mathbf{R}\in {\Omega}}{\text{max}}~~p(\mathbf{B}, \mathbf{R} | \mathbf{O}) \propto p(\mathbf{O} | \mathbf{B}, \mathbf{R}) \cdot p(\mathbf{B}) \cdotp(\mathbf{R}),
\end{equation}
where $\mathbf{O}\in \mathbb{R}^{h \times w}$, $\mathbf{B}\in \mathbb{R}^{h \times w}$, and $\mathbf{R}\in \mathbb{R}^{h \times w}$ denote the observed rainy image, rain-free image, and rain streaks, respectively. $p(\mathbf{B}, \mathbf{R} | \mathbf{O})$ is the posterior probability and $p(\mathbf{O} | \mathbf{B}, \mathbf{R})$ is the likelihood function. $\Omega :=\left\{\mathbf{B}, \mathbf{R} | 0 \leq \mathbf{B}_{i}, \mathbf{R}_{i} \leq \mathbf{O}_{i}, \forall i \in[1, M \times N]\right\}$ is the solution space. Generally, the MAP problem can be equivalently reformulated as the following energy minimization problem~\cite{liubilevel}:
\begin{equation}\label{map}
\underset{\mathbf{B}, \mathbf{R}\in {\Omega}}{\text{min}}~~f\left (\mathbf{O}, \mathbf{B}, \mathbf{R} \right )+\Psi(\mathbf{B})+\Phi(\mathbf{R}),
\end{equation}
where the first term $f\left (\mathbf{O}, \mathbf{B}, \mathbf{R} \right )$ represents the fidelity term measuring the discrepancy between the input image $\mathbf{O}$  and the recovered image ${\mathbf{B}}$. The two regularization terms $\Psi(\mathbf{B})$ and $\Phi(\mathbf{R})$ model image priors on $\mathbf{B}$ and $\mathbf{R}$. Since single image rain removal is an ill-posed inverse problem, the priors play important roles in constraining solution space and enforcing desired property of the output~\cite{zhangkai}.

Various methods have been proposed for designing the forms of all terms involved in
(\ref{map}). By using certain optimization algorithms, generally including an iterative process, the recovery image can then be obtained~\cite{zhangkai}. We introduce representative works presented along this line as follows.

Fu \emph{et al.}\cite{fu} utilized morphological component analysis (MCA) to formulate rain removal as an image decomposition problem. Specifically, a rainy image was divided into LFP and HFP with a bilateral filter, and the derained result was obtained by merging the LFP and the rain-free component. Here the component was achieved by performing dictionary learning and sparse coding on the HFP. For more accurate HFP, Chen \emph{et al.}~\cite{chen_visual2} exploited sparse representation and then separated rain streaks from the HFP by exploiting a hybrid feature set, including histogram of oriented gradients, depth of field, and Eigen color. Similarly, Kang \emph{et al.}~\cite{kang,kang2} exploited histogram of oriented gradients (HOGs) features of rain streaks to cluster into rain and non-rain dictionary. 

To remove rain and snow for single image, Wang \emph{et al.}~\cite{wang} designed a 3-layer hierarchical scheme.
With a guided filter, the authors obtained the HFP consisting of rain/snow and image details, and then decomposed it into rain/snow-free parts and rain/snow-affected parts via dictionary learning and three classifications of dictionary atoms. In the end, with the sensitivity of variance of color image (SVCC) map and the combination of rain/snow detection and the guided filter, the useful image details could be extracted.

Afterwards, Sun \emph{et al.}~\cite{sun} intended to exploit the structural similarity of image bases for single image rain removal. By focusing on basis selection and incorporating the strategy of incremental dictionary learning, the idea is robust against rain patterns and can preserve image information well.

To finely separate rain layer and rain-removed image layer, Luo \emph{et al.}~\cite{dsc} proposed a dictionary learning based single image rain removal method. The main idea is to sparsely approximate the patches of the two layers by high discriminative codes over a learned dictionary with strong mutual exclusivity.
To better remove more rain streaks and preserve background layer, Li \emph{et al.}~\cite{lp} introduced GMM based patch prior to accommodate multiple orientations and scales of rain streaks.


For the progressive separation of rain streaks from background details, Zhu \emph{et al.}~\cite{zhu} modeled  three regularization terms in various aspects: integrating local and nonlocal sparsity via a centralized sparse representation, measuring derivation of gradients from the estimated rain direction by analyzing the gradient statistics, and measuring the visual similarity between image patches and rain patches to filter the rain layer. 

Very recently, Gu \emph{et al.}\cite{gu} proposed a joint convolutional analysis and synthesis (JCAS) sparse representation model, where  image large-scale structures were approximated by analysis sparse representation (ASR) and image fine-scale textures were described by synthesis sparse representation (SSR).
The complementary property of ASR and SSR made the proposed JCAS able to effectively extract image texture layer without oversmoothing the background layer.

Considering the challenge to establish effective regularization priors and optimize the objective function in (\ref{map}), Mu \emph{et al.}~\cite{liubilevel} introduced an unrolling strategy to incorporate data-dependent network architectures into the established iterations, i.e., a learning bilevel layer priors method to jointly investigate the learnable feasibility and optimality of rain streaks removal problem. This is a beneficial attempt to integrate both model-driven and data-driven methodologies for the deraining task.

\subsubsection{\textbf{Deep learning based methods}}
Eigen \emph{et al.}~\cite{eig} first utilized CNN to remove dirt and water droplets adhered to a glass window or camera lens. However, the method fails to handle relatively large/dense raindrops and dynamic rain streaks, and produces blurry outputs. In order to deal with substantial presence of raindrops, Qian \emph{et al.}~\cite{qian} designed an attentive generative network. The basic idea is to inject visual attention into the generative and discriminative networks. Here the generative network focuses on raindrop regions and their surroundings, and the discriminative network mainly assesses the local consistency of restored regions.

To especially deal with single image rain streak removal, Fu \emph{et al.}~\cite{clear} first designed a CNN based DerainNet, which automatically learnt the nonlinear mapping function between clean and rainy image details from data.
To improve the restoration quality, the authors additionally introduced image processing domain knowledge. Motivated by great success of deep residual network (ResNet)~\cite{he_res}, Fu \emph{et al.}~\cite{ddn} further proposed a deep detail network (DDN) to reduce the mapping range from input to output and then to make the learning process significantly easier. Again, Fan and Fu \emph{et al.}~\cite{fan_resi} proposed a residual-guided feature fusion network (ResGuideNet), where a coarse to fine estimation of negative residual was progressively obtained as network goes deeper.

Instead of relying on image decomposition framework like \cite{clear, ddn}, Zhang \emph{et al.}~\cite{zhang_gan} proposed a conditional generative adversarial networks (GAN) for single image deraining which incorporated quantitative, visual, and discriminative performance into objective function. Since a single network may not learn all patterns in training samples, the authors \cite{did} further presented a density-aware image deraining method using a multistream dense network (DID-MDN). By integrating a residual-aware classifier process, DID-MDN can adaptively determine the rain-density information (heavy/medium/light).

Recently, Yang \emph{et al.}~\cite{jorder} reformulated the atmospheric process of rain as a new model, expressed as:
\begin{equation}
\mathbf{O}=\alpha\left(\mathbf{B}+\sum_{t=1}^{s} \widetilde{\mathbf{S}}_{t} \mathbf{R}\right)+(1-\alpha) \mathbf{A},
\end{equation}
where $\mathbf{R}$ denotes the locations of individually visible rain streaks. Each $\widetilde{\mathbf{S}}_{t}$ is a rain streak layer with the same direction and $s$ is the maximum number of layers. $\mathbf{A}$ is the global atmospheric light and $\alpha$ is the atmospheric transmission.
Based on the generative model, the authors developed a multi-task architecture that successively learnt binary rain streak map, appearance of rain streaks, and clean background. By utilizing a RNN and a contextualized dilated network~\cite{dilt}, the method can remove rain streak and rain accumulation iteratively and progressively, even in the presence of heavy rain. For better deraining performance, the authors further proposed an enhanced version--JORDER$\_$E, which included an extra detail preserving step~\cite{tpami}.

Similarly, Li \emph{et al.}\cite{rescan} proposed a recurrent squeeze-and-excitation (SE) based context aggregation network (CAN) for single image rain removal, where SE block assigned different alpha-values to various rain streak layers and CAN acquired large receptive field and better fit the rain removal task.

Existing deep learning methods usually treated network as an encapsulated end-to-end mapping module without deepening into the rationality and superiority towards more effective rain streaks removal~\cite{kernel,pan}. Li \emph{et al.}~\cite{nledn} proposed a non-locally enhanced encoder-decoder network to efficiently learn increasingly abstract feature representation for more accurate rain streaks and then finely preserve image details.

As seen, the constructed  deep network  structures become more and more complicated, making network designing hardly reproducible and attainable to many beginners in this area. To alleviate this issue, Ren \emph{et al.}\cite{pre} presented a simple and effective progressive recurrent deraining network (PReNet) by repeatedly unfolding a shallow ResNet with a recurrent layer.

A practical issue for data-driven single image rain removal methods is the requirement of synthetic rainy/clean image pairs, which cannot sufficiently cover wider range of rain streak patterns in real rainy image such as rain shape, direction and intensity. In addition, there are no public benchmarks for quantitative comparisons on real rainy images, which makes current evaluation less objective. To handle these problems, Wang \emph{et al.}~\cite{spa} semi-automatically constructed a large-scale dataset of rainy/clean image pairs that covers a wide range of natural rain scenes, and proposed a spatial attentive network (SPANet) to remove rain streaks in a local-to-global manner.

As we know, the main problem in recent data-driven single image rain removal methods is that they generally need to pre-collect sufficient supervised samples, which is time-consuming and cumbersome. Besides, most of these methods are trained on synthetic samples, making themselves less able to well generalize to real test samples with complicated rain distribution. To alleviate these problems, Wei \emph{et al.}\cite{semi} adopted DDN as the backbone (supervised part) and regularized rain layer with GMM to feed unsupervised rainy images. In this semi-supervised manner, the method ameliorates the hard-to-collect-training-sample and overfitting-to-training-sample issues.


\subsection{A Comprehensive Repository for Rain Removal}
To facilitate an easy usage and performance reproduction/comparison of current rain removal methods for general users, we build a repository for current research development of rain removal\footnote{\url{https://github.com/hongwang01/Video-and-Single-Image-Deraining}}. Specifically, this repository includes direct links to 74 rain removal papers, source codes of 9 methods for video rain removal and 20 ones for single image rain removal, 19 related project pages, 6 synthetic datasets and 4 real ones, and 4 commonly used image quality metrics as well as their computation codes including peak-signal-to-noise ratio (PSNR)~\cite{psnr}, structure similarity (SSIM)~\cite{ssim}, visual quality (VIF)~\cite{vif}, and feature similarity (FSIM) \cite{fsim}. The state-of-the-art performance for the rain removal problems can thus be easily obtained by general users. All our experiments were readily implemented by using this repository.
%
\begin{figure*}[htb]
    \vspace{-2mm}
\centering
        \includegraphics[scale=0.52]{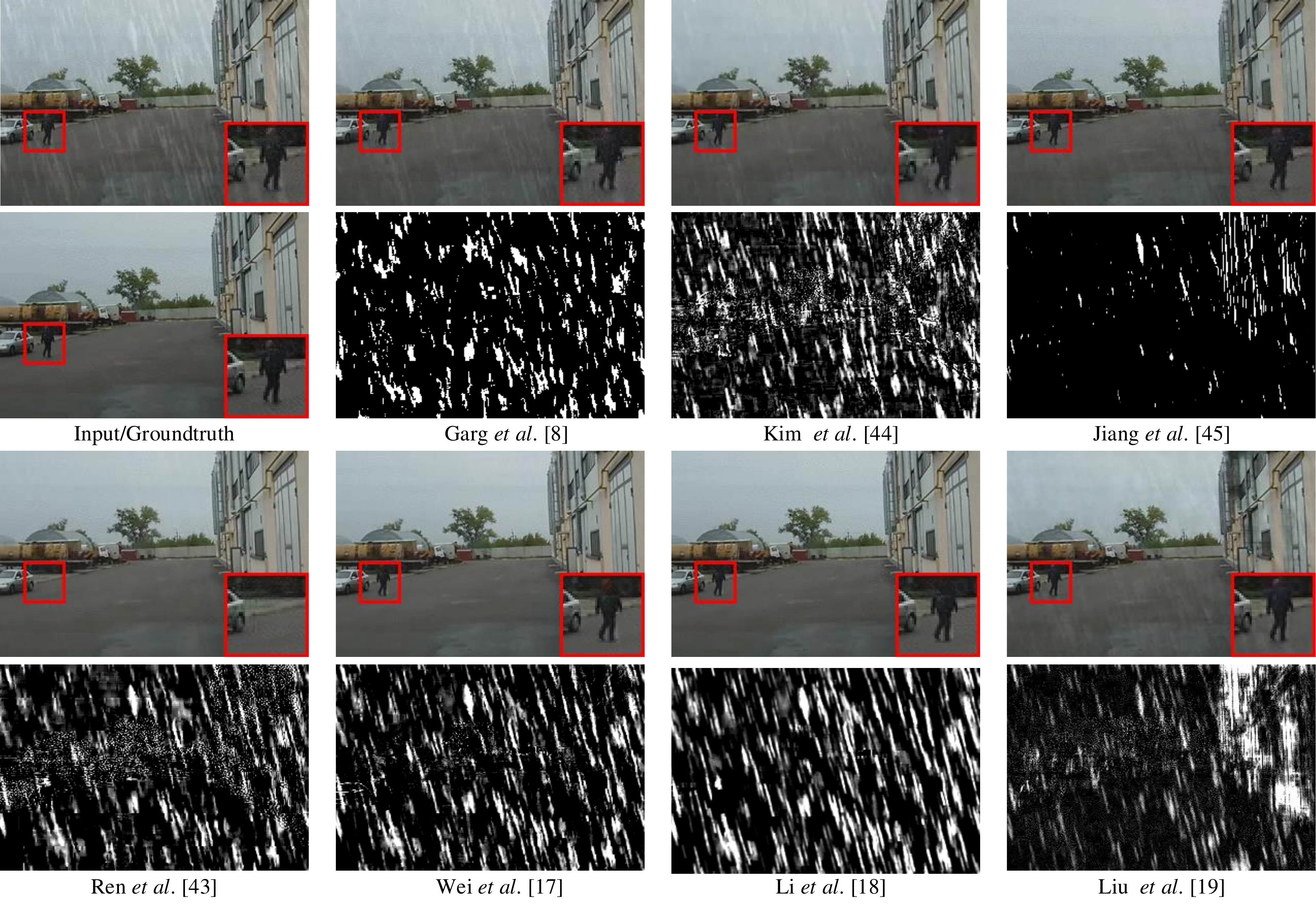}
        \vspace*{-4mm}
    \caption{The first column: input rainy frame (upper) and groundtruth (lower). From the second to the eighth columns: derained frames and extracted rain layers by 7 competing methods.}
    \label{park}\vspace{-3mm}
\end{figure*}
\begin{figure*}[htb!]
    \vspace{-2mm}
\centering
        \includegraphics[scale=0.505]{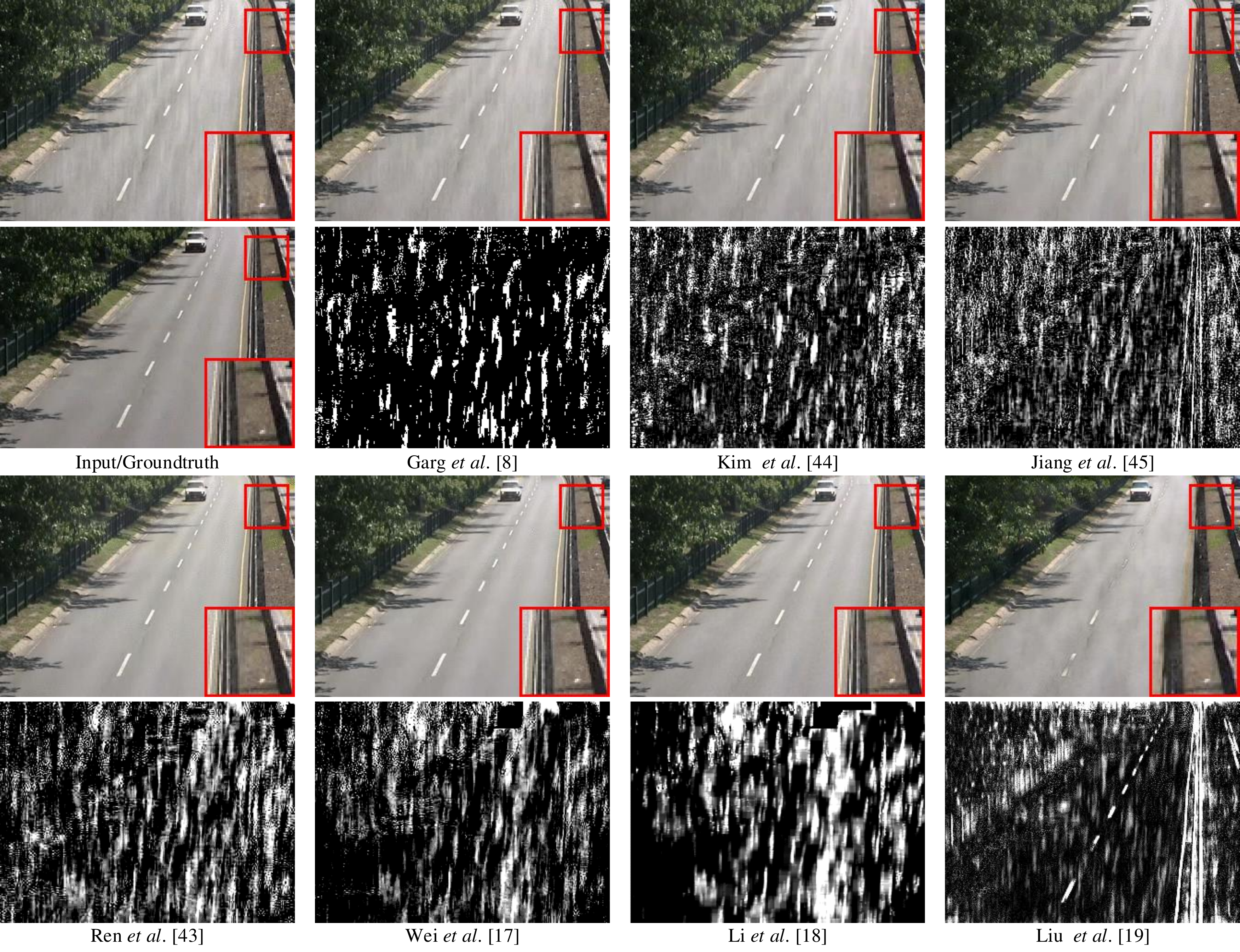}
        \vspace*{-4mm}
    \caption{The first column: input rainy frame with synthetic heavy rain (upper) and groundtruth (lower). From the second to the eighth columns: derained frames and extracted rain layers by 7 competing methods.}
    \label{highway}\vspace{-4mm}
\end{figure*}

\begin{figure*}[htb]
\vspace{-2mm}
  \centering
        \includegraphics[scale=0.448]{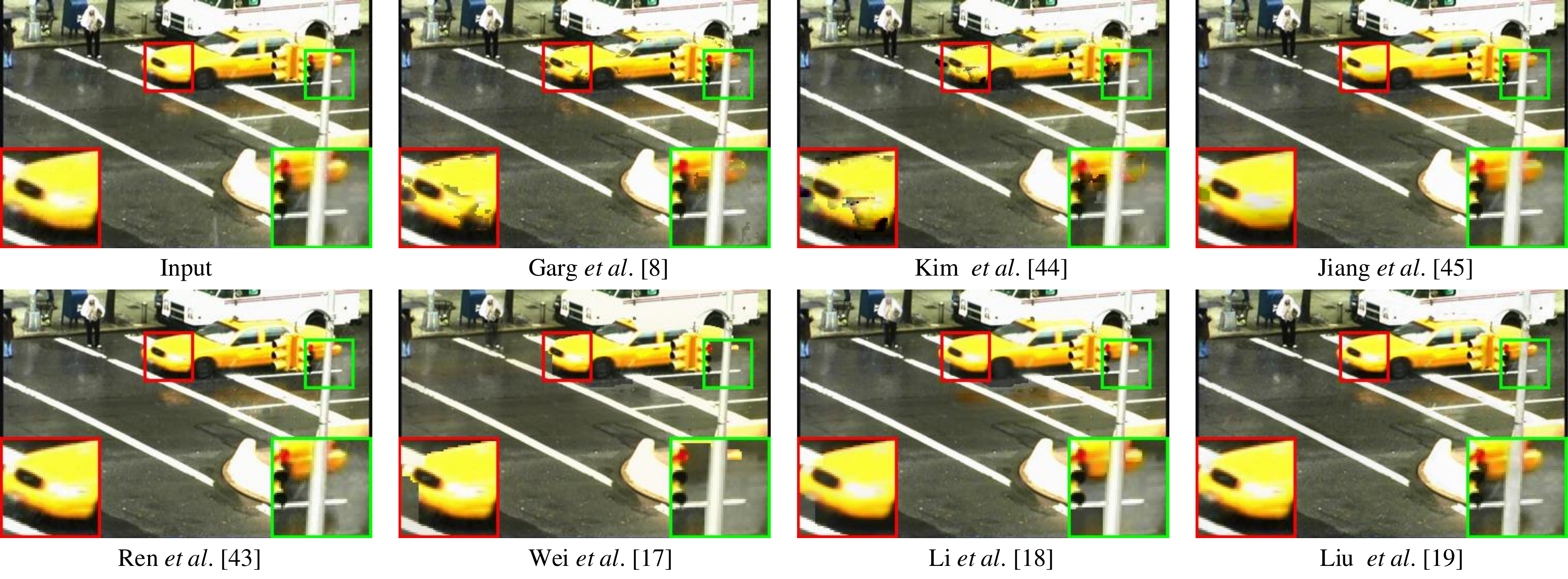}
        \vspace*{-4mm}
        \caption{
        Rain removal performance of all competing methods on a real video with complex moving objects.}
    \label{comp}\vspace{-1mm}
\end{figure*}

\begin{figure*}[htb]
    \vspace{-2mm}
   \centering
        \includegraphics[scale=0.448]{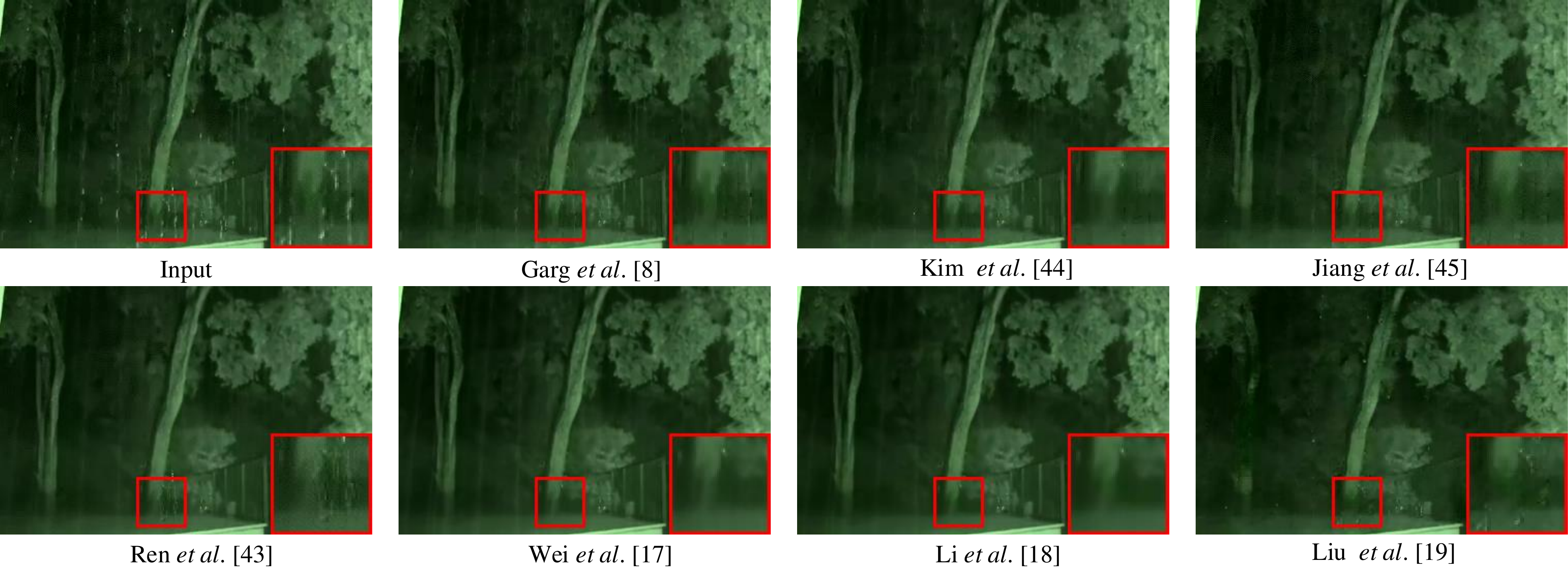}
        \vspace*{-4mm}
    \caption{
    Rain removal performance of all competing methods on a real video at night.}
    \label{night}\vspace{-4mm}
\end{figure*}
\section{Experiments and Analysis}

In this section, we compare the performance of different competing deraining methods for video and a single image. The implementation environment is Windows10, Matlab (R2018b), PyTorch (version 1.0.1)~\cite{pytorch}, and Tensorflow (version 1.12.0) with an Intel (R) Core(TM) i7-8700K at 3.70GHZ, 32GM RAM, and two Nvidia GeForce GTX 1080Ti GPUs.

\subsection{Video Deraining Experiments}
In this section, we evaluate the video deraining performance of the recent state-of-the-art methods on synthetic and real benchmark datasets. These methods include Garg \emph{et al.}~\cite{gk_vision},
Kim \emph{et al.}~\cite{kim},
Jiang \emph{et al.}~\cite{jiang},
Ren \emph{et al.}~\cite{ren},
Wei \emph{et al.}~\cite{wei},
Li \emph{et al.}~\cite{csc},
and Liu \emph{et al.}~\cite{liujia}.

\subsubsection{\textbf{Synthetic Data}}
Here we utilize the dataset released by ~\cite{csc}. The paper chose two videos from CDNET database~\cite{condet}, containing varying forms of moving objects and background scenes, and added different types of rain streaks under black background of these videos, varying from tiny drizzling to heavy rain storm and vertical rain to slash line. For synthetic data, since the rain-free groundtruth videos are available, we can compare all competing methods both visually and quantitatively. Four typical metrics for video have been employed, including PSNR, SSIM, VIF, and FSIM.

Fig. \ref{park} illustrates the deraining performance of all compared methods on videos with usual rain. As displayed in the first row, the rain removal results show that Garg \emph{et al.}'s, Kim \emph{et al.}'s, Jiang \emph{et al.}'s, and Liu \emph{et al.}'s methods do not finely detect rain streaks, and Ren \emph{et al.}'s method improperly removes moving objects and rain streaks. The corresponding rain layers provided in the second row depict that apart from Li \emph{et al.}'s method which can preserve texture details well, the rain layers extracted by the other methods contain different degrees of background information.

We also evaluate all competing methods under heavy rain scenario as shown in Fig. \ref{highway}. The rain removal results displayed in the first row indicate that Garg \emph{et al.}'s, Kim \emph{et al.}'s, Jiang \emph{et al.}'s, and Liu \emph{et al.}'s methods do not well detect heavy rain streaks. In comparison with Wei \emph{et al.}'s method, which treats rain streaks as aggregation of noise rather than natural streamline, Li \emph{et al.}'s method presents natural rain patterns and has a better visual effect.
\begin{table}[t]
\centering
\caption{Performance comparisons of all competing video rain removal methods in synthetic rain.}

\begin{tabular}{@{}C{1.2cm}@{}|@{}C{0.9cm}@{}@{}C{0.9cm}@{}@{}C{0.9cm}@{}@{}C{0.9cm}|C{0.9cm}@{}@{}C{0.9cm}@{}@{}C{0.9cm}@{}@{}C{0.9cm}@{}}
  \hline
\Xhline{1.2pt}
  Datasets & \multicolumn{4}{|c|}{Fig. \ref{park}} & \multicolumn{4}{|c@{}}{Fig. \ref{highway}} \\
\Xhline{0.5pt}
  Metrics & PSNR & VIF & FSIM & SSIM & PSNR & VIF & FSIM & SSIM \\
\Xhline{1.2pt}
  Input & 28.22 & 0.637 & 0.935 & 0.927 & 23.82 & 0.766 & 0.970 & 0.929\\
\Xhline{0.5pt}
  Garg\cite{gk_vision} &29.83 & 0.661 & 0.955 & 0.946 & 24.15 & 0.611 & 0.960 &0.911 \\
\Xhline{0.5pt}
  Kim \cite{kim} & 30.44 & 0.602 & 0.958 & 0.952 & 22.39 & 0.526 & 0.932 &0.886 \\
\Xhline{0.5pt}
  Jiang \cite{jiang}& 31.93 & 0.745 & 0.971& 0.974 & 24.32& 0.713 &0.966 & 0.938\\
\Xhline{0.5pt}
  Ren\cite{ren} & 28.26& 0.685 &0.970 &0.962 &23.52 &0.681 &0.966 & 0.927 \\
\Xhline{0.5pt}
  Wei\cite{wei} & 29.76& 0.830 &0.992 &0.988 &24.47 &0.779 &\textbf{0.980} &0.951 \\
\Xhline{0.5pt}
  Li\cite{csc} &\textbf{33.89} &\textbf{0.865} &0.992 &\textbf{0.992} &\textbf{25.37}& \textbf{0.790} &\textbf{0.980} &\textbf{0.957}\\
\Xhline{0.5pt}
    Liu\cite{liujia}& 27.56 &0.626 &\textbf{0.995} &0.941 &22.19& 0.555& 0.946 &0.895 \\
\Xhline{1.2pt}
\end{tabular}
\label{t1}\vspace{-4mm}
\end{table}

The quantitative comparisons are listed in Table \ref{t1}, which shows that among these competing methods, Li \emph{et al.}'s method achieves a relatively better performance in terms of used quality metrics. Note that the performance of Liu \emph{et al}'s method is not very satisfactory since there is an evident bias between the training data and our testing cases. The overfitting issue inevitably occurs.

\begin{figure*}[htb]
\vspace{-2mm}
        \includegraphics[scale=0.23]{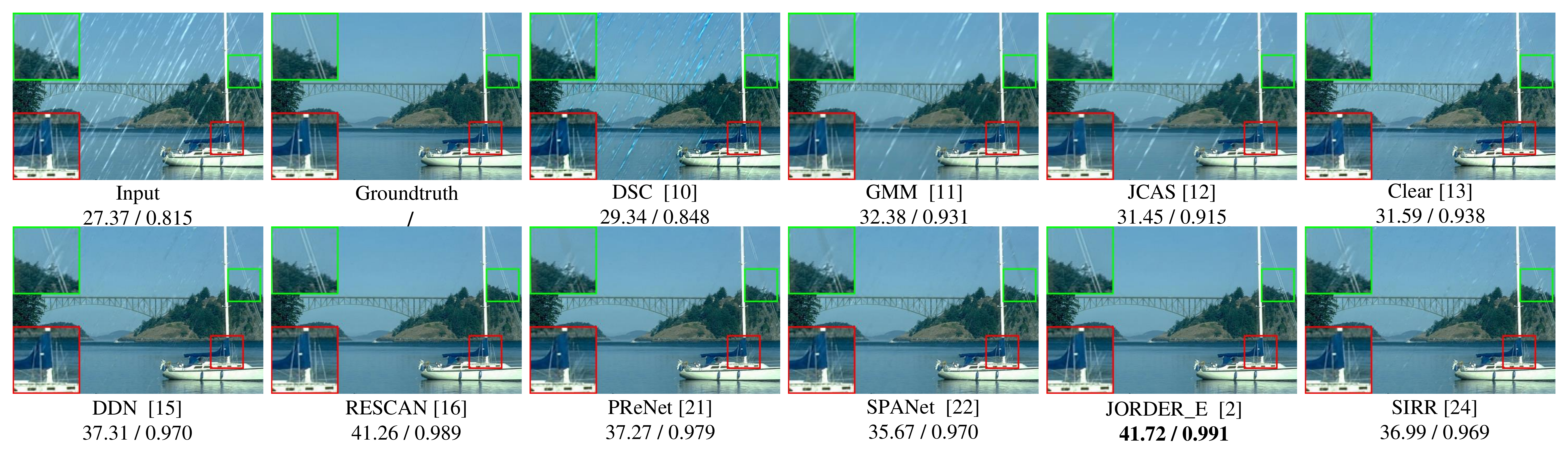}
        \vspace*{-4mm}
    \caption{Rain removal performance of all competing methods on a synthetic test image from Rain100L. PSNR/SSIM results are included below the corresponding recovery image for reference.}
    \label{L}
\end{figure*}
\begin{figure*}[htb]
\vspace{-2mm}
\centering
        \includegraphics[scale=0.23]{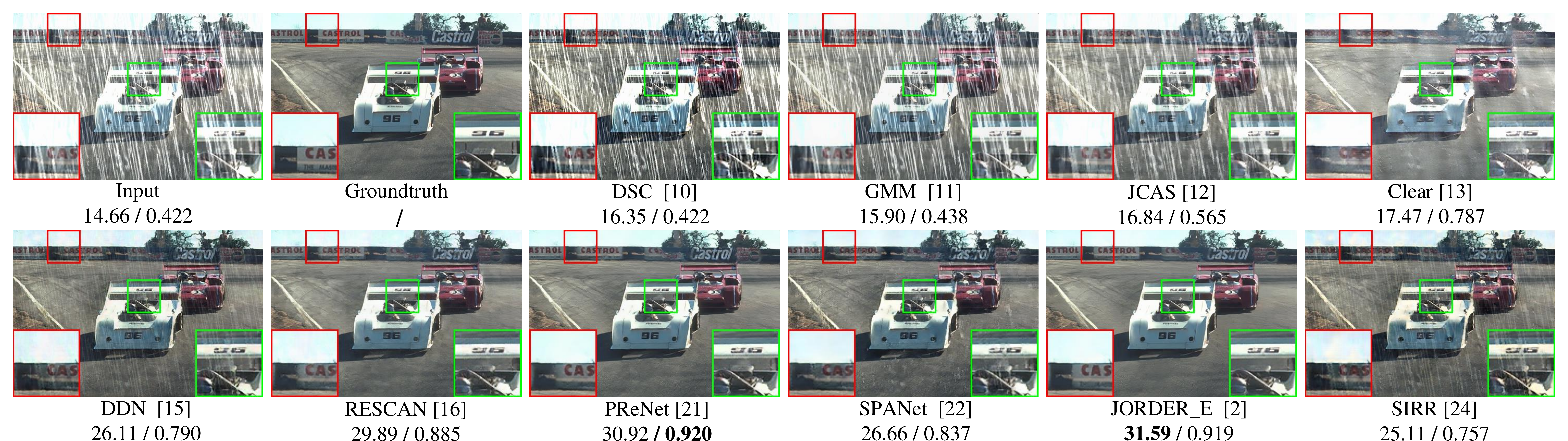}
        \vspace*{-3mm}
    \caption{Rain removal performance of competing methods on a synthetic test image from Rain100H. PSNR/SSIM results are included below the corresponding recovery image for reference.}
    \label{H}
\end{figure*}

\subsubsection{\textbf{Real-World Data}}
We then show the rain streak removal results on real videos. As we have no groundtruth knowledge in this case, we only provide the visual effect comparisons.

Fig. \ref{comp} presents the deraining results on a video with complex moving objects, including walking pedestrian and moving vehicles, which is captured by surveillance systems on street. It is seen that Garg \emph{et al.}'s, Kim \emph{et al.}'s, Jiang \emph{et al.}'s, and Wei \emph{et al.}'s methods cause different degrees of artifacts at the location of the moving car. Comparatively, Li \emph{et al.}'s method performs relatively well in this complicated scenario.

Fig. \ref{night} displays the rain removal performance on a real video obtained at night. Comparatively, Wei \emph{et al.}'s and Li \emph{et al.}'s methods can better detect all rain streaks.
\subsection{Single Image Deraining Experiments}
In this section, we evaluate the single image deraining performance of the recent state-of-the-art methods, including typical model-driven methods:
Luo \emph{et al.}~\cite{dsc}
(denoted as DSC),
Li \emph{et al.}~\cite{lp} (denoted as GMM),
and Gu \emph{et al.}~\cite{gu} (denoted as JCAS),
and representative data-driven methods: Fu \emph{et al.}~\cite{clear} (denoted as Clear),
Fu \emph{et al.}~\cite{ddn} (denoted as DDN),
Li \emph{et al.}~\cite{rescan} (denoted as RESCAN),
and Ren \emph{et al.}~\cite{pre} (denoted as PReNet),
Wang \emph{et al.}~\cite{spa} (denoted as SPANet),
Yang \emph{et al.}~\cite{tpami} (denoted as JORDER\_E),
and semi-supervised method:
Wei \emph{et al.}~\cite{semi} (denoted as SIRR).

\subsubsection{\textbf{Synthetic Data}}
For synthetic data, we utilized four frequently-used benchmark datasets: Rain1400 synthesized by Fu~\emph{et al.} \cite{ddn}, Rain12 provided by Li~\emph{et al.}\cite{lp}, Rain100L and Rain100H provided by Yang \emph{et al.}~\cite{jorder}. Specifically, Rain1400 includes 14000 rainy images synthesized from 1000 clean images with 14 kinds of different rain streak orientations and magnitudes. Among these images, 900 clean images (12600 rainy images) are chosen for training and 100 clean images (1400 rainy images) are selected as testing samples. Rain12 consists of 12 rainy/clean image pairs. Rain100L is selected from BSD200~\cite{bsd} with only one type of rain streaks, which consists of 200 image pairs for training and 100 image pairs for testing. Compared with Rain100L, Rain100H with five types of streak directions is more challenging, which contains 1800 image pairs for training and 100 image pairs for testing. As for SIRR, we use the real 147 rainy images released by Wei \emph{et al.}\cite{semi} as unsupervised training data. Since Rain12 has few samples, like\cite{pre}, we directly adopt the trained model on Rain100L to do an evaluation on Rain12.

As the groundtruth in synthetic datasets is available, we will evaluate all competing methods by two commonly used metrics PSNR and SSIM.
Since human visual system is sensitive to Y channel in YCbCr space, we utilize the luminance channel to compute all quantitative results.

Fig. \ref{L} shows the visual and quantitative comparisons of rain streak removal results for one synthesized rainy image from Rain100L. As displayed, three model-driven methods: DSC, GMM, and JCAS, leave many rain streaks in the recovered image. Especially, JCAS tends to oversmooth the background details. It implies that model prior is not sufficient enough to convey complex rain streak shapes in synthetic dataset. Compared with these conventional model-driven methods, six data-driven methods, Clear, DDN, RESCAN, PReNet, SPANet, and JORDER\_E,  have the ability to more completely remove the rain streaks. However, they damage the image content and lose detail information to a certain extent. Although SIRR focus on domain adaption, it fails to remove most rain streaks. This can be explained by the fact that there exists an obvious difference in distribution between Rain100L and real rainy images.

\begin{figure*}[htb]
\vspace{-1mm}
\centering
        \includegraphics[scale=0.176]{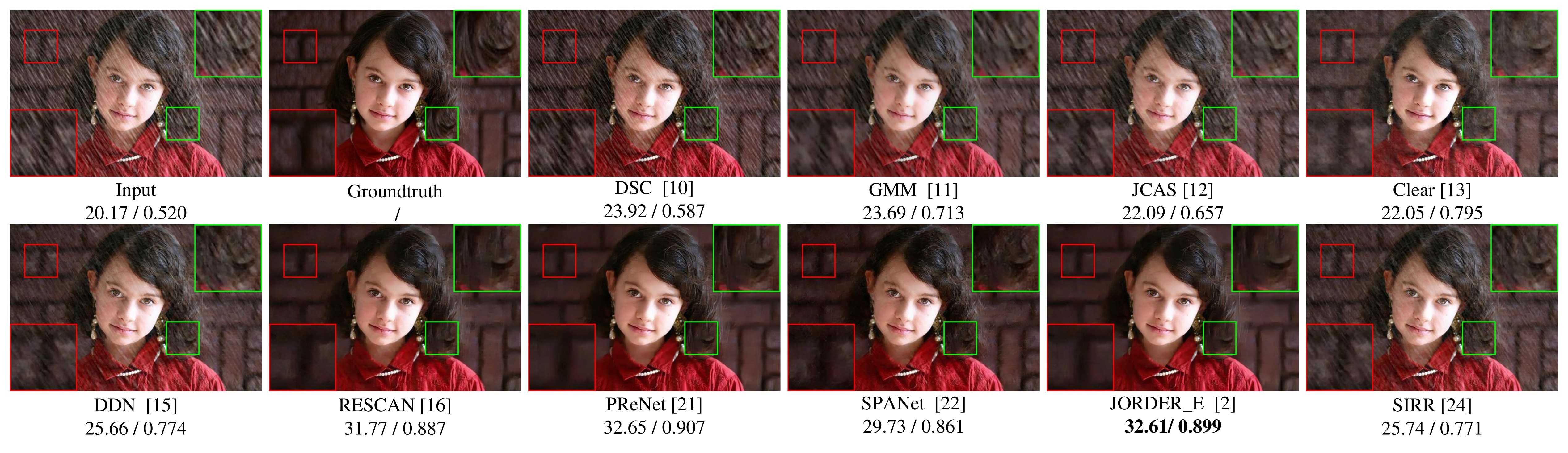}
        \vspace*{-4mm}
    \caption{Rain removal performance of competing methods on a synthetic test image from Rain1400. PSNR/SSIM results are included below the corresponding recovery image for reference.}
    \label{1400}
\end{figure*}
\begin{figure*}[htb]
\centering
        \includegraphics[scale=0.23]{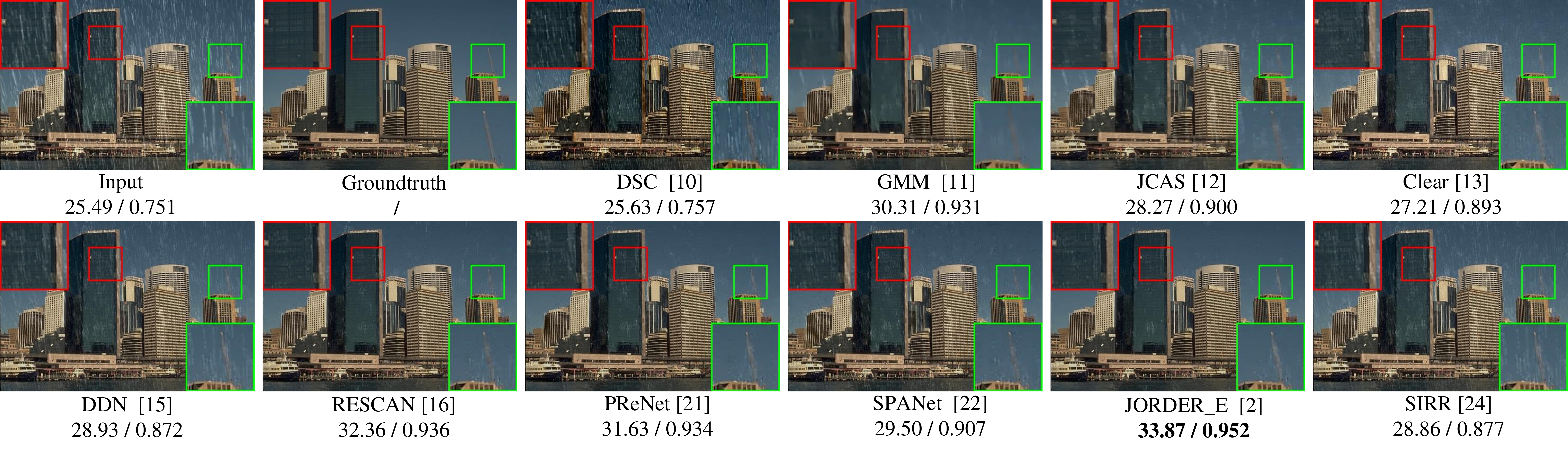}
        \vspace*{-4mm}
    \caption{Rain removal performance of competing methods on a synthetic test image from Rain12. PSNR/SSIM results are included below the corresponding recovery image for reference.}
    \label{12}
\end{figure*}
We further evaluate these competing methods on Rain100H. As shown in Fig. \ref{H}, due to complicated rain patterns in heavy rain cases, the rain detection capability of most competing methods is weakened. By observing zoomed red boxes, we can find that for almost all competing methods, the rain removal results are not very satisfactory when rain streaks and the image background merge with each other. More rational and insightful understanding for intrinsic imaging process of rain streaks is still required to be further discovered and utilized~\cite{jorder}.

We additionally made an evaluation based on Rain1400 and Rain12 with different rain patterns as presented in Fig.~\ref{1400} and Fig.~\ref{12}. From these, we can easily understand that generally the data-driven methods can achieve better rain removal effect than model-driven methods. However, due to the overfitting-to-training-samples issue, these deep learning methods make derained results lack of some image details.
\begin{table}[t]
\centering
\caption{PSNR and SSIM comparisons on four synthetic benchmark datasets.}
\setlength{\tabcolsep}{3.1mm}{
\begin{tabular}{@{}C{2.0cm}@{}|@{}C{0.82cm}@{}C{0.82cm}@{}|@{}C{0.82cm}@{}C{0.82cm}@{}|@{}C{0.82cm}@{}C{0.82cm}@{}|@{}C{0.82cm}@{}C{0.82cm}@{}}
\hline
\Xhline{1.2pt}
  Datasets & \multicolumn{2}{|c|}{Rain100L} & \multicolumn{2}{|c|}{Rain100H} & \multicolumn{2}{|c|}{Rain1400} & \multicolumn{2}{|c@{}}{Rain12} \\
\Xhline{0.5pt}
  Metrics & PSNR & SSIM & PSNR & SSIM & PSNR & SSIM & PSNR & SSIM \\
\Xhline{1.2pt}
  Input & 26.90 & 0.838 & 13.56 & 0.371 & 25.24 & 0.810 & 30.14 & 0.856\\
\Xhline{0.5pt}
  DSC\cite{dsc}& 27.34 & 0.849 & 13.77 & 0.320 & 27.88 &0.839 & 30.07 &0.866\\
\Xhline{0.5pt}
  GMM\cite{lp}&29.05 &0.872 & 15.23 &0.450 &27.78 & 0.859 & 32.14 & 0.916 \\
\Xhline{0.5pt}
  JCAS\cite{gu}& 28.54 & 0.852 & 14.62 & 0.451 &26.20 & 0.847 & 33.10 &0.931\\
\Xhline{0.5pt}
  Clear\cite{clear}&30.24 & 0.934 & 15.33 & 0.742 & 26.21& 0.895 & 31.24 & 0.935\\
\Xhline{0.5pt}
  DDN\cite{ddn}& 32.38 & 0.926 & 22.85 & 0.725  & 28.45 & 0.889 & 34.04 & 0.933\\
\Xhline{0.5pt}
  RESCAN\cite{rescan} & 38.52& 0.981 &29.62 & 0.872  &32.03& 0.931 &36.43&0.952 \\
\Xhline{0.5pt}
  PReNet\cite{pre}& 37.45& 0.979 &\textbf{{30.11}} & 0.905 & 32.55 & \textbf{0.946}& 36.66& 0.961\\
\Xhline{0.5pt}
  SPANet\cite{spa} & 34.46 & 0.962 &25.11 & 0.833 & 29.76 & 0.908 & 34.63& 0.943\\
  \Xhline{0.5pt}
  JORDER\_E\cite{tpami}&{{\textbf{38.61}}} &{\textbf{0.982}} & 30.04 &{\textbf{0.906}} & \textbf{32.68} & 0.943 &\textbf{36.69}&\textbf{0.962} \\
  \Xhline{0.5pt}
  SIRR\cite{semi} & 32.37 & 0.926 & 22.47 & 0.716 & 28.44 & 0.889 & 34.02& 0.935\\
\Xhline{1.2pt}
\end{tabular}}
\label{t2}\vspace{-5mm}
\end{table}

Table~\ref{t2} demonstrates quantitative results of all competing methods on synthetic datasets. From the table, we can conclude that due to the strong nonlinear fitting ability of deep networks, the rain removal effect of most data-driven methods is evidently superior than those of model-driven methods. Besides, compared with the backbone network--DDN, SIRR hardly obtains any performance gain on these datasets. This can be explained by the fact that the usage of real unsupervised training samples makes the data distribution deviate from synthetic datasets.

\subsubsection{\textbf{Real-World Data}}
For real application, what we really care about is the deraining ability of all competing methods on real rainy images. Here we will give a fair evaluation based on two real-world datasets: the one with 147 rainy images released by Wei \emph{et al.}~\cite{semi}, called Internet-Data, and the other with 1000 image pairs collected by Wang \emph{et al.}~\cite{spa}, called SPA-Data. Note that as Internet-Data has no groundtruth, we can only provide visual comparison.

\begin{figure*}[htb]
\centering
        \includegraphics[scale=0.1465]{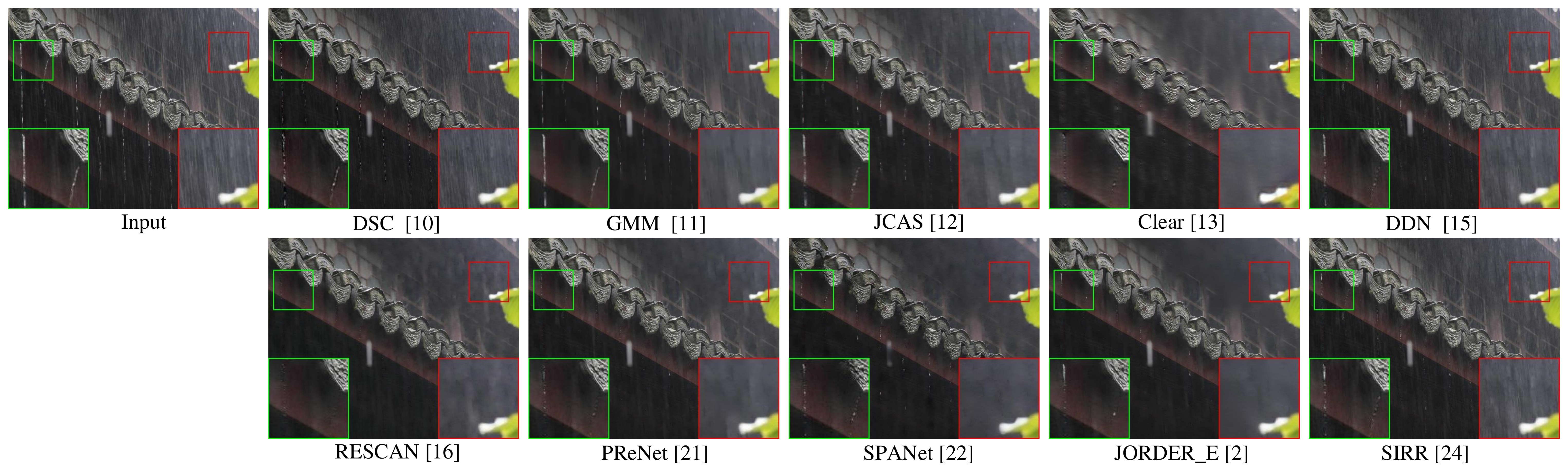}
        \vspace*{-4mm}
    \caption{Rain removal performance of different methods on a real rainy image from \cite{semi}.}
    \label{wei}
\end{figure*}
\begin{figure*}[htb]
\centering
        \includegraphics[scale=0.21]{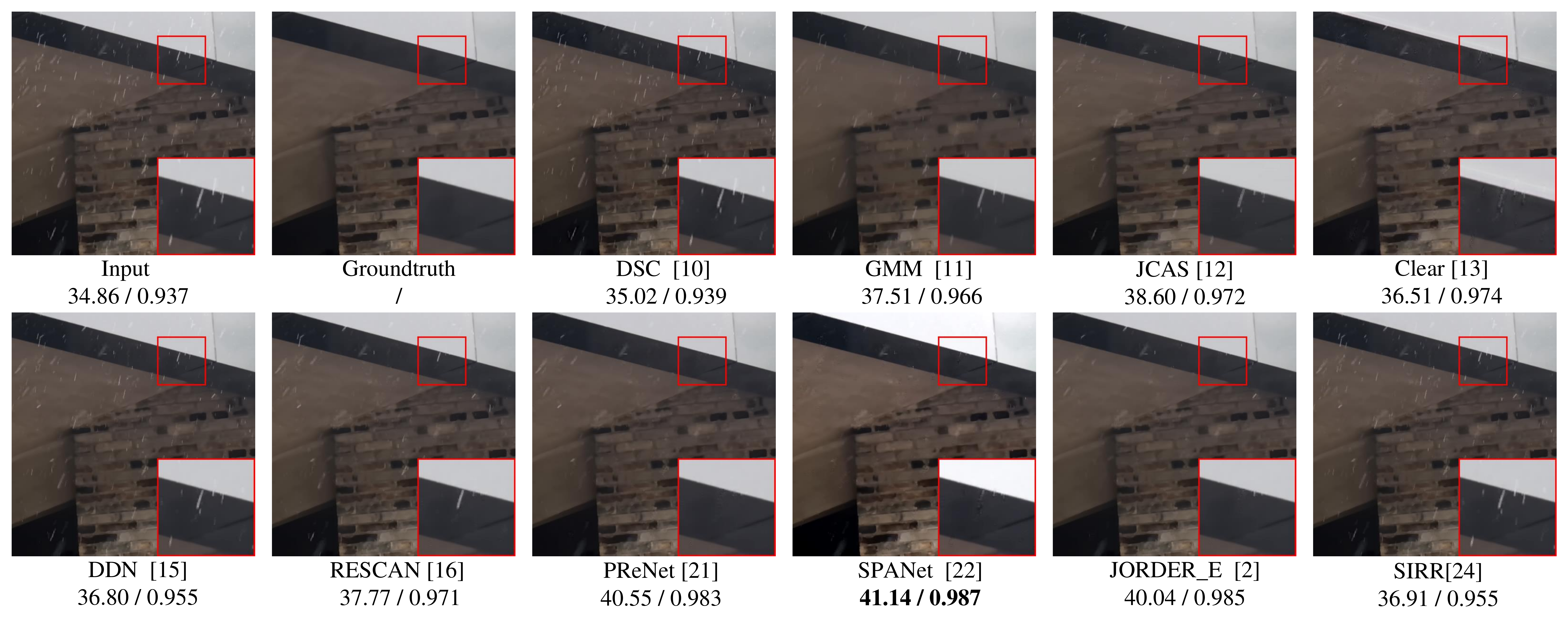}
        \vspace*{-4mm}
    \caption{Rain removal performance of different competing methods on a real rainy image from \cite{spa}. PSNR/SSIM results are included below the corresponding recovery image for reference.}
    \label{wang}\vspace{-2mm}
\end{figure*}

Fig.~\ref{wei} demonstrates a hard sample with various rain densities selected from Internet-Data. As seen, almost all competing methods cannot completely remove rain streaks and perfectly clear up rain accumulation effect. Even though PReNet, RESCAN, and JORDER\_E achieve significant deraining performance on synthetic datasets, they oversmooth the background information to some extent. This can be interpreted as that for model-driven methods, the priors they adopt have not comprehensively covered the complicated distribution of real rain, and for data-driven methods, they tend to learn specific rain patterns in synthesized data while cannot properly generalize to real test samples with diverse rain types.

Further, we utilize SPA-Data to more objectively analyze the generalization ability of all competing methods as displayed in Fig. \ref{wang} and Table~\ref{t4}. These comparisons tell us that in this case, the model-driven method JCAS with meaningful priors even performs better than some data-driven works, i.e., DDN and RESCAN. It is worth mentioning that although the rain removal performance of SPANet on synthesized datasets with imprecise rain mask is not very satisfying, it obtains an outstanding generalization ability on the real dataset with easily extracted rain mask. Additionally, compared with DDN, SIRR accomplishes a better transfer learning effect, which benefit from the unsupervised module.

%
\begin{table}[t]
\centering
\caption{PSNR and SSIM comparisons on SPA-Data\cite{spa}.}\vspace{-1mm}
\begin{tabular}{@{}C{1.8cm}@{}|@{}C{1.2cm}@{}@{}C{1.2cm}@{}|@{}C{2.1cm}@{}|@{}C{1.2cm}@{}@{}C{1.2cm}@{}@{}}
  \hline
\Xhline{1.2pt}
 Methods & PSNR  & SSIM & Methods & PSNR  & SSIM\\
\Xhline{1.2pt}
 Input & 34.15  & 0.927  & RESCAN\cite{rescan}  & 34.70  &0.938 \\
\Xhline{0.5pt}
DSC\cite{dsc} & 34.95  & 0.942 &PReNet\cite{pre} & 35.08  &0.942  \\
\Xhline{0.5pt}
GMM\cite{lp}  &34.30  & 0.943 & SPANet\cite{spa}  &\textbf{35.24} & \textbf{0.945} \\
\Xhline{0.5pt}
JCAS\cite{gu}  & 34.95 &\textbf{0.945} & JORDER\_E\cite{tpami} & 34.34 &0.936 \\
\Xhline{0.5pt}
Clear\cite{clear} & 32.66  & 0.942 &   SIRR \cite{semi} & 34.85   & 0.936 \\
\Xhline{0.5pt}
DDN\cite{ddn}  & 34.70   &0.934 & / & / &/\\
\Xhline{1.2pt}
\end{tabular}
\label{t4}\vspace{-4mm}
\end{table}

\section{Conclusions and Future Works}
In this paper, we have presented a comprehensive survey on the rain removal methods for video and a single image in the past few years. Both conventional model-driven and latest data-driven methodologies raised for the deraining task have been thoroughly introduced. Recent representative state-of-the-art algorithms have been implemented on both synthetic and real benchmark datasets, and the deraining performance, especially the generalization capability have been empirically compared and quantitatively analyzed. Especially, to make general users easily attain rain removal resources, we release a repository, including direct links to 74 rain removal papers, source codes of 9 methods for video rain removal and 20 ones for single image rain removal, 19 related project pages, 6 synthetic datasets and 4 real ones, and 4 commonly used image quality metrics. We believe this repository should be beneficial to further prompt the further advancement of this meaningful research issue.
Here we want to summarize some limitations still existing in current deraining methods as follows:
\begin{enumerate}
\item Due to the intrinsic overlapping between rain streaks and background texture patterns, most of deraining methods tend to more or less remove texture details in rain-free regions, thus resulting in oversmoothing effect in the recovered background.
\item  As aforementioned, the imaging process of rain in natural scenes is very complex~\cite{benchmark,jorder,pan}. However, the rain model widely used in  most existing methods has not sufficiently describe such intrinsic mechanism, like the mist/fog effect formed by rain streak accumulation.
\item Although current model-driven methods try to portray complex rain streaks by diverse well-designed priors, they are only applicable to specific patterns instead of irregular distribution in real rainy images. Another obvious drawback is that the optimization algorithms employed by these methods generally involve many iterations of computation, causing their inefficiency in real scenarios~\cite{dsc,lp,gu,ren,liubilevel}.
 \item Most data-driven methods require a great deal of training samples, which is time-consuming and cumbersome to collect~\cite{ddn,jorder,spa}. And they generally have unsatisfactory generalization capability because of the overfitting-to-training-sample issue. Besides, the designed networks are always like black boxes with less interpretability and few insights~\cite{liu,liutip}.
 \item For the video deraining task, most model-driven methods cannot directly apply to streaming video data~\cite{ren, wei,csc} in real-time. Meanwhile, the deep learning methods need a large amount of supervised videos, which exhibits high computational complexity in training stage, and they cannot guarantee favorable rain removal performance especially in complex scenes~\cite{chen3,liujia}.
\end{enumerate}

The rain removal for video and a single image is thus still an open and worthy to be further investigated problem. Based on our evaluation and research experience, we also try to present the following remarks to illustrate some meaningful future research directions along this line:
\begin{enumerate}
  \item  Due to the diversity and the complexity of real rain, a meaningful scope is to skillfully combine model-driven and data-driven methodologies into a unique framework to make it possess both superiority of two learning manners. A hopeful direction is the deep unroll strategy, which might conduct networks with both better interpretability and generalization ability~\cite{liutip}.

  \item To deal with the hard-to-collect-training-example and overfitting-to-training-example issues, semi-supervised/unsupervised learning as well as domain adaption and transfer learning regimes should be necessary to be explored to transfer the learned knowledge from limited training cases to wider range of diverse testing scenarios~\cite{semi,un_gan}.

 \item To better serve real applications, we should emphasize efficiency and real-time requirement. Especially for videos, it is necessary to construct online rain removal techniques which meet three crucial properties: persistence (process steaming video data in real time), low time and space complexity, universality (available to complex video scenes). Similarly, fast test speed for a single image is also required.

 \item Generally, deraining is served as a pre-processing step for certain subsequent computer vision tasks. It is critical to develop task-specific deraining algorithm~\cite{benchmark}.

   \item Due to the fact that rain is closely related to other weather conditions, such as haze and snow~\cite{csc,jorder}, multi-task learning by accumulating multiple sample sources collected in bad weathers for performance enhancement may also be worth exploring in future research.
\end{enumerate}

\ifCLASSOPTIONcaptionsoff
  \newpage
\fi

\bibliography{bib_tip}

\end{document}